# MODE MATCHING IN THE FIELD SPECKLE PATTERN AFTER GRADED-INDEX FIBERS

**Alexander Volyar, Mikhail Bretsko, Yana Akimova, Yuriy Egorov**
*V.I. Vernadsky Crimean Federal University,*
*Vernadsky Prospect, 4, Simferopol, 295007, Russia.*

**Abstract**
We have developed and implemented a new technique of parallel information transmission through a multimode graded-index fiber in real time based on the approach of intensity moments. To do this, it was necessary to measure the spectra of the mode amplitudes and phases in the Hermite-Gauss and Laguerre-Gauss basis using the transfer matrix of the fiber. At the same time, informational entropy and the degree of spatial coherence were measured.

A set of fiber eigenmodes subjected to random perturbations forms a chaotic speckle-structure that can be treated either as the inevitable noise of the information transmission system, or as a set of independent parallel information channels. The first case implies suppressing or reducing speckle structure, while the second one is aimed at using a mode array for parallel big data transmission, in particular, optical images. This is only a small circle of the current problems of modern photonics requiring quick solutions [1]. The development and implementation of artificial neural networks [2,3] enable one employing pattern recognition techniques via big data processing. In particular, the authors of Ref. [4,5] employ the linear transformations (the transfer matrix) methods that links an array of image points at a spatial light modulator (SLM) in front of the fiber input to points of the speckle structure at the fiber output (or the charge-coupled device (CCD)). The general principles of seeing through a turbid medium were developed and implemented in Ref.[4]. The authors of Ref.[5.6] associated the input pattern points with a group of degenerate fiber modes. For example, in Ref. [6] for transmission of the mode composition, a sequence of individual modes for each basis polarization is first transmitted through the fiber. At the fiber output, each mode is sequentially displayed on the phase mask of the SLM device. Further, all phase masks are added together. By varying the mode phases and amplitudes, the maximum signal power is achieved that corresponds to the optimal mode matching. Such a complex sequence of operations can be appreciably simplified by using spiral beams [7,8] for parallel signal transmission while the eigenmodes spectra are measured at the fiber output. Indeed, the spiral beams due to strong mode matching are capable of transporting complex images in free space, and at the multimode fiber input it is rather simple to describe them in terms of degenerate eigenmodes. However, when propagating along the fiber, the spiral beam properties in the eigen modes are lost. To measure the mode spectra (amplitudes and phases) in various bases, the intensity moments technique was developed and implemented allowing direct measurements of complex mode compositions [9-11]. Thus, our letter is the first demonstration of analyzing the mode speckle structure of a multimode graded-index fiber, measuring mode spectra, information entropy and degree of spatial coherence as well as the speckle structure matching that restores the initial mode

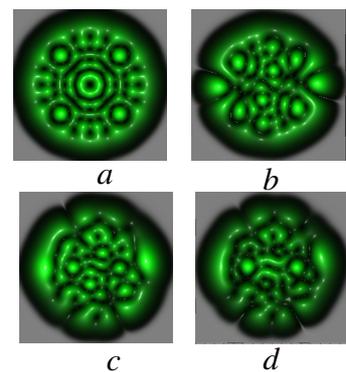

Fig.1 Computer simulation of the speckle pattern evolution in fiber with $V \approx 9.1$, $\varepsilon = 1$, $a = 10$ : (a) the only $LG_{4,+1}$ mode and $a = 0$, (b) the same but $a = 10$; (c) $LG_{4,+1} + LG_{1,-3} + LG_{3,+2}$ modes; (d) 21 modes with the same amplitudes

pattern.

For simplicity, we restricted ourselves to a multimode graded-index fiber model with an unbounded parabolic profile (see, i.e. [12,13]).

$$n^2 = n_1^2 \left[1 - 2\Delta(r/w)^2\right], 1 \leq 2\Delta(r/w)^2, \quad (1)$$

where $w = (2/\kappa)^{1/2}(n_1 n_2)^{-1/4}$ stands for an effective fiber core radius, $\Delta = n_2/(2n_1)$, $\kappa$ is a wavenumber, $n_1$ is a refractive index at the fiber axis, $n_2$ is a core heterogeneity coefficient. The eigen vortex modes in a circularly polarized basis form a uniformly polarized pattern [14] in the form

$$\mathbf{E}^{(+,-)}_{\pm l,n} = \mathbf{e}_\pm R^l \exp\left(\pm i l \varphi - i \tilde{\beta} z\right) L_n^{(l)}\left(VR^2\right) \exp\left(-VR^2/2\right), \quad (2)$$

где $L_n^{(l)}(VR^2)$ is Laguerre polynomial, $V = \kappa w n_1 \sqrt{2\Delta}$, $R = r/w$, while the scalar propagation constant is $\tilde{\beta} = (V/w\sqrt{2\Delta})\sqrt{1 - 4\Delta(2n+l-1)/V}$. (3)

The spin – orbit coupling [15] in the vortex mode introduces an additional correction to the propagation constant $\beta = \tilde{\beta} + \delta\beta$ so that $\delta\beta \sim 1 m^{-1}$ in comparison with $\tilde{\beta} \sim 10^7 m^{-1}$. We will assume that the fiber material has birefringence $\Delta n = n_x - n_y$ with an optical axis directed perpendicular to the $z$ fiber axis. If $\kappa \Delta n \gg \delta\beta$, then the spin – orbit coupling is suppressed [16] and birefringence $\kappa \Delta n$ makes the main contribution to the polarization correction. In this case, the circularly polarized components of field (2) turn into linearly polarized $E_\pm \to E_{x,y}$ ones, and the propagation constant becomes $\beta_{x,y} = \tilde{\beta} + \Delta\beta_{x,y}$. Note that the eigenmodes of the parabolic fiber are both the Laguerre – Gauss modes and the Hermite – Gauss modes where , that are linked as [11]]

$$LG_{n,\pm l}(\mathbf{r}, z|\varepsilon) = \frac{(-1)^n}{2^{2n+3l/2}} \sum_{k=0}^{2n+l} (\mp i)^k \lambda_k \varepsilon_k HG_{2n+l-k,k}(\mathbf{r}, z), \quad (4)$$

$$HG_{N-k,k} = H_{V-k}\left(\sqrt{2}VX\right) H_{N-k}\left(\sqrt{2}VY\right) e^{-X^2-Y^2-i\beta_{n,l} z}, \quad (5)$$

where $(X,Y) = (x/w, y/w)$, $N = 2n+l$, $\lambda_k = (-2)^k P_k^{(n+l-k,n-k)}(0)$ $H_{n,m}(x)$ is Hermite polynomial. $P_k^{(n+l-k,n-k)}$ is Jacobi polynomial and $\varepsilon_k = 1 - \varepsilon(-1)^k \sin(k\pi/2) \exp(i(1+aP))$, is a perturbation. In the nonperturbed case they present a phase matched mode group being one of the spiral beam types that preserve its the propagation-invariant property up to scale transformations. In our model, we neglected intermodal coupling when mode propagating through the fiber. Besides, strong birefringence allows us to confine ourselves to measuring the mode spectrum in each linearly polarization component. In our model, we restrict ourselves to random phase perturbations of each HG fiber mode in (4) with a random number generator on each element of a finer length, so that a random function $-\pi > P_k > \pi$, $\varepsilon$ and $a$ is a regular and random amplitude perturbation, respectively, in Eq.(5). Fig. 1 shows the result of computer simulation of speckle structure in our multimode fiber model with a waveguide number $V = 9.1$. A regular intensity distribution (Fig.1a) corresponds to a mode $n=4, l=1, \varepsilon=1$ in absence of a random perturbation. The random perturbation of this mode in Fig. 1b only partially violates its regularity. Growing the number of modes in Fig. 1c, d leads to complete randomization of the speckle structure. In the general case, the HG modes in the wave mixture (4) can be treated as independent parallel information transmission channels. In our

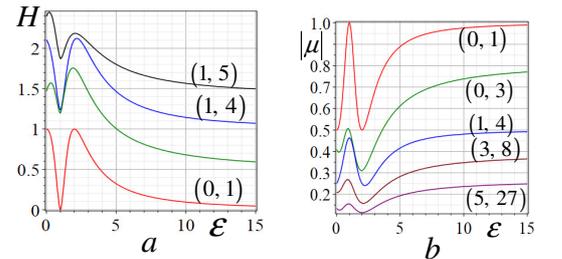

Fig.2 (a) Informational entropy $H$ and (b) degree of special coherence $\mu$ as a function of regular $\varepsilon$ perturbation for different numbers $(n, l)$, $A_{n,l} = 1$

experiments, we will use compositions of degenerate LP modes in the form

$$E_{\pm}(\mathbf{r},z) = \sum_{n,l=0}^{n_{max},l_{max}} A_{n,l} \frac{(-1)^n}{2^{2n+3l/2}} \sum_{k=0}^{2n+l} C_k^{(\pm)} HG_{2n+l-k,k}(\mathbf{r},z), \quad (6)$$

where $A_{n,l}$ and $C$ are amplitudes of the degenerate LG and HG fiber mode groups in eq, (4). The fiber informational compositions in terms of normalized mode amplitudes are described by the modulus of spatial coherence degree [17]

$$|\mu| = \sum_{n,l=0}^{n_{max},l_{max}} |A_{n,l}|^4 \sum_{k=0}^{2n+l} |C_k^{(\pm)}|^4 / \left[ \sum_{n,l=0}^{n_{max},l_{max}} |A_{n,l}|^2 \sum_{k=0}^{2n+l} |C_k^{(\pm)}|^2 \right]^2 \quad (7)$$

and the information entropy [10]

$$H = -\sum_{n,l=0}^{n_{max},l_{max}} |A_{n,l}|^2 \sum_{k=0}^{2n+l} |C_k^{(\pm)}|^2 \log |C_k^{(\pm)}|^{2 \cdot 2}, \quad (8)$$

Both statistical characteristics $H$ and $|\mu|$ for single $LG_{n,l}$ mode groups shown in Fig.2 have common features: for large perturbations $\varepsilon \gg 1$, they tend to a finite values $H \to H_0(n,l), |\mu| \to \mu_0(n,l)$ that depends only on the topological charge $l$ and the radial number $n$. They can be treated as the statistical invariants of the fiber speckle pattern. From the physical point of view, this means that the growing of the wavefront dislocations number runs out. At small perturbations $0 < \varepsilon < 3$, oscillations of both entropy $H$ and coherence $|\mu|$ are observed that corresponds to the mode reconstructions. Concluding the discussion of the model, it is important to note that since the LG mode group in Eq.(4) with the same sum of numbers $2n+l$ is degenerate in each polarized component, degeneracy vanishes due to random perturbations. Each perturbed LG mode within the $2n+l$ group is excited in the fiber in the form of a set of independent HG modest with random phases.

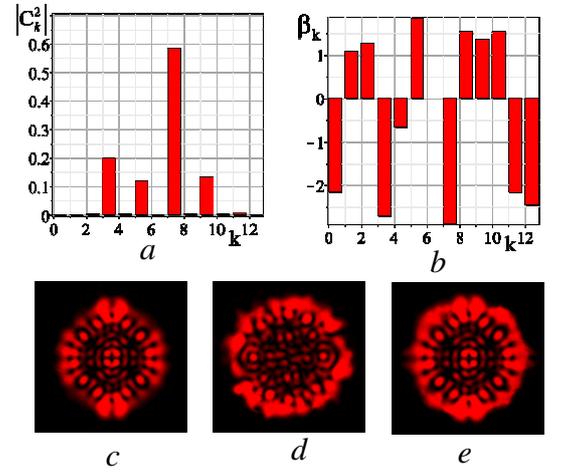

Fig.3 Mode spectrum (a,b) and restoring initial mode pattern (c-e): {a) squared amplitudes $|C_k|^2$, (b) random mode phases $\beta_k$; (c) initial mode pattern, (d) speckle structure, (e) restoring mode pattern. The correlation degree is $\eta = 0.95$, a regular perturbation is $\varepsilon = 2$, the initial mode numbers of LG mode are $n = 5, l = 2$.

In the experiment, we performed two types of measurements. First of all, we measured the spectrum of modes: the spectrum of the squared amplitudes and phases of each HG mode. Based on mode spectra, at first the dependences of information entropy $H(\varepsilon)$ and the degree of spatial coherence $|\mu(\varepsilon)|$ as functions of the regular perturbation $\varepsilon$ were calculated using Eq, (7) and (8) (solid curves in Fig.2). Then we measured experimental points placed in Fig.2a and b. As can be seen from the figure, the experimental points are in good agreement with the theoretical curves. Secondly, we measured the spectrum of HG modes in a composition of LG modes subject to random perturbations, matched the mode phases and amplitudes in a speckle pattern, and restored the initial mode intensity distribution.

A multimode graded-index fiber was computer simulated by a spatial light modulator (Thorlabs EXULUS-4K1), while the mode spectra are calculate by scanning the intensity distribution in the charge-coupled device (CCD Thorlabs DCC1645) and measuring the intensity moments. The graded-index fiber has the waveguide parameter $V \approx 14,2$ The fiber can maintain $N \approx 50$ LG modes with maximum mode numbers $l_{max} = 10, n_{max} = 10$ and has a strong birefringence.

The process of measuring the spectra of LG and HG modes and the experimental set-up were discussed in detail in Ref. [10-12,18]. In general, a mode composition in the fiber input includes the whole range of LG modes that can be realized in the fiber, with LG modes having different amplitudes $A_{n,l}$, while degenerate LG modes decay into a spectrum of HG modes with different random phases $\beta_k$. In order to match the phases and amplitudes of the modes at the fiber output, two mode spectra have to be measured: the HG mode spectrum, which gives information about the phases $\beta_k$ and amplitudes $C_k$ of the HG modes, and the LG mode spectrum, which gives additional information about the amplitudes $A_{n,l}$ of the degenerate LG modes. The simplest case was the measurement of the spectrum of the HG modes in a separate perturbed LG mode. To do this, it was sufficiently to use the method described in detail in Ref, [11]. In order to determine the mode spectra, we compiled a set of linear equations

$$J_{p,q} = \frac{1}{J_{00}} \iint_{\mathbb{R}^2} F_{p,q} \Im_{n,\ell}(x,y) dx dy \tag{8}$$

for the squared amplitudes $|C_{2n+l-k}|^2$ and cross terms containing phase differences $\beta_{k,j} = \beta_k - \beta_j$ in the intensity distribution $\Im_{n,\ell}(x,y)$, where $F_{p,q} = H_p(\sqrt{2}x)H_q(\sqrt{2}y)$ is the moments function, $p,q = 0,1,2,...$. Equation (8) can be written in the matrix form $\mathbf{J} = M\,\mathbf{C}$, where $\mathbf{J}$ denotes the column-vector of the intensity moments $J_{p,q}$, and $\mathbf{C}$ is the column-vector of the squared amplitudes and cross terms of the intensity distribution, then the inverse transformation $\mathbf{C} = M^{-1}\mathbf{J}$ specifies immediately the mode spectra, while $M^{-1}$ is the transfer matrix of the fiber. To plot the spectrum of mode phases, we need to know the phase of only one mode (say $\beta_k = 0$) Figure 3a,b shows an example of the spectrum of HG modes subjected to random perturbations in a degenerate group of LG modes with the mode numbers $n = 5, l = 2$ and the regular perturbation $\varepsilon = 2$. The experimental points in Fig.2a, b represent the dependence of entropy $H(\varepsilon)$ and degree of spatial coherence $|\mu|$ on a regular perturbation $\varepsilon$ under the condition of random perturbations. The scatter of points relative to theoretical curves does not exceed 3%. Each point corresponds to a separate spectrum of squared amplitudes. The spectra of HG mode amplitudes and phases together with the known value of the regular perturbation $\varepsilon = 2$, allows us to create an algorithm for the mode matching the intensity distribution that restores the initial mode pattern shown in Fig. 3c-e with the correlation degree $\eta = 0.95$ of the initial and output mode patterns. The main contribution to a small decorrelation is made by the framing in Fig.3e caused by the lateral modes in the spectrum of Fig.3a,b. To measure the HG mode spectrum of a complex wave composition, the HG mode spectrum is first measured outlined in Fig. 4a,b, .The electric field composed of a combination of 50 degenerate LG modes (or 50 eigen HG modes) is transmitted through a graded-index birefringent fiber with $V \approx 14.2$, subject to random perturbations. This spectrum allows one to calculate only the degree of coherence and informational entropy of the beam at the fiber output. However, to match the mode amplitudes and phases that form the speckle picture and restore the initial pattern of the mode composition, it is still necessary to measure the spectrum of LG modes. Measuring the spectrum $A_{n,l}$ of the LG modes (not shown in the figure), we used the method det, we employed the technique detailed in Ref. [18], and using our developed computer algorithm we matched the amplitudes and phases of the HG modes. The results obtained demonstrate Fig.4c-e. The intricate regular intensity distribution in Fig. 4c turns under the random perturbations into a typical speckle pattern in Fig. 4b. The phase and amplitude matching restores the initial mode composition in Fig. 4e with the degree of correlation $\eta = 0.93$. It is worth noting that despite the fact that mode coupling is not taken into account in our fiber model, two spectral measurement of the LG and HG modes makes it possible to restore the initial mode pattern when modes are coupled. In fact, each HG mode in the speckle pattern represents an independent information channel, and the whole process can be treated as a parallel information transfer.

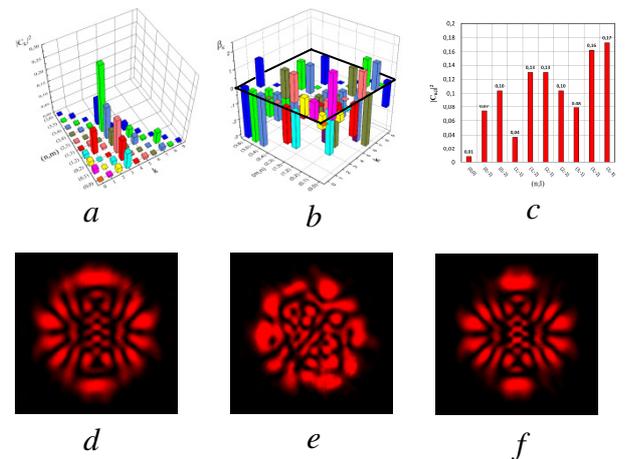

Fig.4 Mode spectrum (a,b,c) and restoring process (d-f) of the $N = 50$ HG mode composition with $l_{max} = 10$, $n_{max} = 10$: (a) squared amplitudes, (b) mode phases; (c) initial mode pattern, (d) speckle structure, (e) restored mode pattern with the correlation degree $\eta = 0.93$.